\documentclass[a4paper,11pt]{article}
\usepackage{jheppub}

\usepackage[english]{babel}
\usepackage{amsmath,amssymb}

\preprint{Imperial-TP-2025-CH-6}

\title{The CFT of Sen's Formulation of Chiral Gauge Fields}

\author[a]{Chris Hull,}
\author[b]{Neil Lambert,}
\affiliation[a]{The Blackett Laboratory, Imperial College London, Prince Consort Road, London, SW7 2AZ, UK}
\affiliation[b]{
Department of Mathematics,
 King's College London, 
 London WC2R 2LS, UK}

\emailAdd{c.hull@imperial.ac.uk, neil.lambert@kcl.ac.uk}

\abstract
{
Sen's action for chiral bosons in 2 dimensions describes two chiral scalars, one of which couples to the physical   metric and one of which couples to a flat  metric. It has a generalisation in which the flat metric is replaced by an arbitrary second   metric and so can be formulated on any curved world-sheet.
 When the two metrics are equal,
 the  theory reduces to   a $\beta \gamma$ system, giving a non-unitary $c=2$
conformal field theory. We argue  that
the relation between this and the theory of   two chiral bosonic scalars of the same chirality
can be viewed as a \lq bosonisation'.
    We show that the standard vertex operators for the chiral scalars are vertex operators and line operators in the Sen formulation and derive the  formulation in the Sen theory of correlation functions in the chiral scalar theory.  The flat space Sen theory can  be coupled to two different world-sheet metrics in such a way that one  scalar couples to one metric and the other to the other metric, so obtaining the general formulation with two metrics.
    
In $d=4k+2$ dimensions,  the bi-metric action for  a $2k$-form gauge field with self-dual field strength reduces, when the two metrics are equal, to a conformal field theory with a $BF$-type action, except that $B$ is a self-dual $d/2$-form and $F$ is a $d/2$-form field strength, $F=dP$. The self-duality of $B$ means that this is not a topological theory but instead represents two self-dual gauge fields.
This has a generalisation to a democratic action for $p$-form gauge fields in any dimension.
 }

\begin{document}
\maketitle

\flushbottom

\section{Introduction}

Sen's   action for $2k$-form gauge fields in $d=4k+2$ dimensions with self-dual field strength  \cite{Sen:2015nph,Sen:2019qit} 
is a theory of the physical  $2k$-form chiral gauge field $A$ plus a second $2k$-form chiral gauge field $C$ which decouples from the physical theory.
The physical   gauge field $A$  couples to the spacetime metric $g$ and the other physical fields, while 
the gauge field $C$  doesn't couple to the spacetime metric $g$ and the other physical fields. Instead, $C$ couples to a flat metric $\eta$ and $G$ is self-dual with respect to  $\eta$.
The interaction with the space-time metric $g$ is governed by a tensor $M$ which was constructed perturbatively in  \cite{Sen:2015nph,Sen:2019qit}.
This theory has been further explored in \cite{Hull:2023dgp,Lambert:2019diy,Andriolo:2020ykk,Barbagallo:2022kbt,Vanichchapongjaroen:2024tkj,Aggarwal:2025fiq,Hull:2025yww, Andriolo:2021gen,Lambert:2023qgs,Mamade:2025vcp}.  

In \cite{Hull:2023dgp}, a generalisation of Sen's action was constructed in which the non-physical gauge-field $C$ couples to an arbitrary second metric 
$\bar g$ instead of the flat metric $\eta$, and its field strength $G$ is self-dual with respect to  $\bar g$.
The theory then describes a physical sector consisting of the gauge field $A$, the spacetime metric $g$ and the other physical fields, together with a shadow sector consisting of the gauge-field $C$ and the second metric $\bar g$. The shadow sector decouples from the physical sector.
This has a number of advantages over Sen's formulation. First, it can be used on any spacetime manifold, not just those admitting a flat metric $\eta$.
Second, it facilitates a geometric construction of the tensor $M$ in closed form (using results from \cite{Andriolo:2020ykk}). Indeed, finding the tensor $M$, which has a complicated dependence on both metrics  $ g$ and $\bar g$ was a  non-trivial part of the the construction of  \cite{Hull:2023dgp}. Next, it has two gauge symmetries with vector parameters, one for which $ g$ is the gauge field  and one for which $\bar g$ is the gauge field. The diffeomorphism symmetry is a diagonal subgroup of these two symmetries. The quantum theory and associated partition function when $g\ne \bar g$ was discussed extensively in \cite{Hull:2025rxy} and a related bi-metric string field theory action is proposed in \cite{Hull:2025mtb}.

On setting the two metrics equal, the theory becomes an interesting conformal field theory in $d=4k+2$ dimensions 
and it is this CFT that is the subject of this paper.
The action is
\begin{equation}
S=\int Q'\wedge dP ,
\end{equation}
where $P$ is a $2k$-form gauge field and $Q'$ is a self-dual $2k+1$-form.
Without the self-duality constraint $Q'=\ast Q'$, this would be a BF topological field theory. However, with the self-duality constraint the theory describes two free self-dual gauge fields.  In $d=2$ dimensions, this is a $\beta\gamma $ system where $\gamma $ has conformal dimension $0$.
We will analyse this system and show that  the equivalent theory of 2 chiral scalars  can be viewed as a bosonisation of this.
We then show how correlation functions of chiral boson vertex operators arise from the correlation functions of vertex operators and line operators in the Sen CFT. We then  discuss the generalisation to higher dimensions.

\section{The Action for Self-Dual Gauge Fields}

\subsection{General dimensions}

The generalisation  
  \cite{Hull:2023dgp}, 
 of Sen's theory \cite{Sen:2015nph,Sen:2019qit}   is a theory with two metrics on the
spacetime: a ``physical'' metric $g_{\mu \nu}$ which couples to all the
physical fields and carries the gravitational degrees of freedom, and a
second metric $\bar{g}_{\mu \nu}$ which doesn't couple to the physical fields.
The degrees of freedom of the theory in $d = 2 q$ dimensions with $q = 2 n +
1$ are a $q - 1$-form field $P$ and a $q$-form field $Q$ which is
$\bar{g}$-self-dual with respect to $\bar{g}_{\mu \nu}$, $Q = \bar{\ast}
Q$.\footnote{The Hodge dual for the physical metric $g_{\mu \nu}$ will be
denoted here by $\ast$ and the Hodge dual for the ``auxiliary'' metric
$\bar{g}_{\mu \nu}$ by $\bar{\ast}$. } The action is
\begin{equation}
  \label{act} S =  \int \left( -\frac{1}{2} dP \wedge \bar{\ast} dP +  Q \wedge
  dP + \frac{1}{4} Q \wedge M (Q) \right) ,
\end{equation}
and Sen's theory is recovered when $\bar{g} = \eta$, the flat Minkowski
metric. Here $M=-\bar\star M$ is a linear map on $q$-forms $Q$ given explicitly in   \cite{Hull:2023dgp}, 
 which
depends on both metrics.\footnote{The $Q,M$ here has been rescaled compared to those in  \cite{Andriolo:2020ykk,Hull:2023dgp}. The $Q$  here is related to the $Q_{old}$ in \cite{Andriolo:2020ykk,Hull:2023dgp} by $Q=Q_{old}/2$. The map $M$ has been rescaled so that $M(Q)=M_{old}(Q_{old})$, so that in 2 dimensions the components $M_{--}=2M_{--old}$. }
 The field equations are   \cite{Hull:2023dgp}, 
\begin{equation}
  \label{feq1} d \left(   \bar{\ast} dP + Q \right) = 0 ,
\end{equation}
\begin{equation}
  \label{feq2} \frac12M + dP = \bar{\ast}  \left(\frac12 M + dP\right) .
\end{equation}
Defining
\begin{equation}
  \label{Giss} G \equiv    \frac 1 2 (dP + \bar{\ast} dP) +  \frac 1 2 Q ,
\end{equation}
which is $\bar{g}$-self-dual
\begin{equation}
  \bar{\ast} G = G ,
\end{equation}
and
\begin{equation}
 \label{Fiss}
  F \equiv  \frac 1 2 Q +  \frac 1 2M (Q) ,
\end{equation}
the field equations (\ref{feq1}),(\ref{feq2}) imply
\begin{equation}
  dG = 0, \qquad dF = 0 .
\end{equation}
There are then $q - 1$-form potentials $A, C$ with $F = dA$, $G = dC$.

The key point is that $M (Q)$ is constructed so that $F$ is $g$-self-dual
\begin{equation}
  \ast F = F .
\end{equation}
This is then a theory of the desired $(q - 1)$-form $A$ with self-dual field
strength $\ast F = F$ and an auxiliary $(q - 1)$-form $C$ whose field strength
is self-dual with respect to the background metric, $\bar{\ast} G = G$. It is
important that the auxiliary field $C$ does not couple to the physical metric
$g_{\mu \nu}$ and the physical field $A$ does not couple to the auxiliary
metric $\bar{g}_{\mu \nu}$.

\subsection{Symmetries and Currents} \label{symcur}

The presence of two metrics $g,\bar g$ leads to two independent gauge symmetries with vector parameters, the $\zeta$-symmetry for which $g$ is the gauge field and under which $\bar g$ is inert and the 
$\chi$-symmetry for which $\bar g$ is the gauge field and under which $ g$ is invariant \cite{Hull:2023dgp}.
The $\zeta$-symmetry acts only on the physical sector, so that $G$ is invariant, 
and it acts as a diffeomorphism on the physical sector (up to on-shell-trivial transformations):
\begin{equation}
  \label{cdiff} \delta G = 0, \quad \delta \bar{g} = 0, \quad \delta g
  =\mathcal{L}_{\zeta} g
  \quad  \delta F \approx \mathcal{L}_{\zeta} F ,
\end{equation}
where $ \delta F \approx \mathcal{L}_{\zeta} F$ indicates that $ \delta F$ is $ \mathcal{L}_{\zeta} F$ on-shell.
The $\chi$-symmetry acts only on the shadow sector, so that $F$ is invariant on-shell, 
and it acts as a diffeomorphism on the shadow sector (up to on-shell-trivial transformations):
\[  \delta \bar{g} =\mathcal{L}_{\chi}  \bar{g}, \quad \quad \delta g = 0 .
\]
\[
  \delta G =  \mathcal{L}_{\chi} G, \quad \delta F  \approx 0 .\]
The transformations of $Q,P$ under these symmetries are given in \cite{Hull:2023dgp}.

The diagonal subgroup with $\zeta=\chi$ acts on all fields and arises from the diffeomorphism symmetry $x^\mu \to x^\mu + \xi^\mu  $ with $\xi^\mu =\tfrac12(\zeta^\mu+\chi^
\mu)$ combined with an on-shell trivial symmetry.

Since there are   two gauge symmetries with gauge fields given by the two metrics, there are two conserved currents 
 defined by the response of the action to a change in the two metrics:
\begin{align}
\Theta_{\mu\nu} = -\frac{2}{\sqrt{-g}} \frac{\delta S}{\delta g^{\mu\nu}},\\
\bar \Theta_{\mu\nu} = -\frac{2}{\sqrt{-\bar g}} \frac{\delta S}{\delta \bar g^{\mu\nu}}	.
\end{align}
Here $\Theta_{\mu\nu} $ can be viewed as the energy-momentum tensor for the physical sector while
$\bar\Theta_{\mu\nu} $ can be viewed as the energy-momentum tensor for the shadow sector.
These are calculated in  Appendix A and found to be
\begin{align}\label{Theta}
\Theta_{\mu\nu} = \frac{1}{(2k)!} 
g^{\rho_1\lambda_1}...g^{\rho_{2k}\lambda_{2k}} F_{\mu\rho_1...\rho_{2k} }F_{\nu\lambda_1...\lambda_{2k} }   .
\end{align}
\begin{align}\label{barTheta}
\bar \Theta_{\mu\nu}\approx -
\frac{1}{(2k)!}\bar g^{\rho_1\lambda_1}...\bar g^{\rho_{2k}\lambda_{2k}} G_{\mu\rho_1...\rho_{2k} }G_{\nu\lambda_1...\lambda_{2k} }    .	
\end{align}
Off-shell, there is a further contribution to $\bar \Theta_{\mu\nu}$ which is given in   Appendix A.
Both of these are traceless and independently conserved on-shell:
 \begin{align}
g^{\mu\nu} \Theta_{\mu\nu}	 &= \bar g^{\mu\nu} \bar\Theta_{\mu\nu}=0,\\
  \nabla^\mu  \Theta_{\mu\nu} 
	 &= \bar \nabla^\mu\bar \Theta_{\mu\nu} =0 .
\end{align}

In the rest of this paper we will mainly  be interested in the case where $g=\bar g$ so $M=0$. This motivates the introduction of a third energy momentum tensor:
\begin{align}
	T_{\mu\nu} & = \bar \Theta_{\mu\nu} +\Theta_{\mu\nu} .
\end{align}
As we show in   Appendix A,  this arises from the diffeomorphism symmetry mentioned above.  
In particular when $g=\bar g$ we find that $g^{\mu\nu}T_{\mu\nu}=0$ and $\nabla^\mu T_{\mu\nu}=0$.

\subsection{Two  Dimensions} \label{2dsub}

Consider the theory in two dimensions on a timelike cylinder or 2d-Minkowski space
with
coordinates $x^{\pm} = \frac{1}{\sqrt{2}}  (x^1 \pm x^0)$, and $\bar{g}$ given
by the Minkowski metric $\bar{g} = \eta$, with line element $ds^2 = 2 dx^+
dx^-$, $\epsilon_{+ -} = 1$, so that for a 1-form with components $V_{\mu}$
\begin{equation}
  (\bar{\ast} V)_{\pm} = \pm V_{\pm} .
\end{equation}
The action (\ref{act}) for a scalar $P$ and 1-form $Q_{\mu}$ which satisfies
$Q = \bar{\ast} Q$ so that $Q_- = 0$ is given by
\begin{equation}
  S = \int d^2 x  (\partial_+ P \partial_- P +  Q_+ \partial_-
  P +  \frac{1}{4} M_{- -} Q_+ Q_+)  ,\label{2dacts}
\end{equation}
with $M (Q)_- = M_{- -} Q_+$. The closed 1-forms $F, G$ with $G = \bar{\ast}
G$ so that $G_- = 0$ and $F = \ast F$ are given by
\begin{equation}
  G_+ =  \frac 1 2 (\partial_+ P + Q_+), \qquad F_+ =  \frac 1 2 Q_+, \qquad F_- = \frac12M_{- -}
  Q_+ .
\end{equation}
The scalar fields $A, C$ with $F = dA$, $G = dC$ then satisfy the self-duality
conditions
\begin{equation}
  \partial_- C = 0, \qquad \partial_- A = M_{- -} \partial_+ A .
\end{equation}
See   \cite{Hull:2023dgp}
 for further discussion. The energy-momentum tensors given above become simply
\begin{align}
\begin{array}{lccrr}
\Theta_{++} = F_+F_+&& \Theta_{--}=0 ,& \hskip.3cm& \Theta_{-+}=0 ,\\
\bar\Theta_{++} =-G_+G_+,& & \bar \Theta_{--}\approx0   ,& \hskip.3cm& \bar\Theta_{-+}=0 ,\\
  T_{++}	  = F_+F_+-G_+G_+,&& T_{--}  \approx0,&\hskip.3cm&T_{-+}=0 .	
\end{array}
\end{align}

\section{Sen-type Action In Two-Dimensional Minkowski Space}

\subsection {The Action}

Consider the 2-dimensional theory (\ref{2dacts}) with $g = \bar{g}$ and hence $M=0$
so that the action is
\begin{equation}
  \label{act222} S = \int \left( \frac{1}{2} dP \wedge \ast dP +  Q \wedge dP
  \right) =   \int Q' \wedge dP ,
\end{equation}
where
\begin{equation}
  Q' = Q + \frac{1}{2} (dP + \ast dP) ,
\end{equation}
is self-dual, $Q' = \ast Q'$. 
The theory is conformally invariant, so in conformal gauge we can take
  $g=\bar g=\eta$ and hence, in the notation of subsection \ref{2dsub},
\begin{equation}
  S = \int d^2 x \, (\partial_+ P \partial_- P +  Q_+ \partial_- P )= \int
  d^2 x \, Q'_+ \partial_- P ,\label{QPact}
\end{equation}
where
\begin{equation}
  Q'_+ = Q_+ + \partial_+ P .
\end{equation}
The field equations are
\begin{equation}
  \partial_- Q'_+ = 0 ,
\end{equation}
\begin{equation}
  \partial_- P = 0 ,
\end{equation}
and on-shell there is a chiral scalar $S$ such that $Q'_+$=$\partial_+ S$ satisfying
$\partial_- S= 0$. Comparing with the last section, it is a theory of two
chiral bosons $A$ and $C$ with field strengths
\begin{equation}
  F_+ =  \frac 1 2 Q_+ =  \frac 1 2 (Q'_+ -  \partial_+ P), \quad G_+ = \frac 1 2 ( Q'_+ + 
  \partial_+ P) =  \frac 1 2 Q_+ + \partial_+ P . \label{FGQP}
\end{equation}
given by
\begin{equation}
  F_+ = \partial_+ A, \quad G_+ = \partial_+ C ,\label{FGAC}
\end{equation}
with
\begin{equation}
  A =\frac 1 2( S-   P), \quad C = \frac 1 2(S+   P) .
\end{equation}

\subsection {The CFT}

The theory with lagrangian
\begin{equation}
  L =  Q'_+ \partial_- P ,
\end{equation}
is often referred to as a $\beta \gamma$ system. Comparing with e.g.
\cite{Polchinski:1998rq,Polchinski:1998rr}:
\begin{equation}
  \beta_+ = Q'_+, \quad \gamma = P, \quad \lambda' = 1 ,
\end{equation}
where $\lambda'$ is the conformal weight of $\beta_+ = Q'_+ .$ 
This is a $c = 2$ chiral CFT corresponding to the stress-energy tensor 
\begin{equation}
  T_{+ +} = -Q'_+ \partial_+ P ,
\end{equation}
 and is non-unitary.  
On the other hand, there is a theory of two chiral bosons
$A, C$ which is a $c = 2$ chiral CFT with stress-energy tensor
\begin{equation}
  T_{+ +} = \partial_+ A \partial_+ A - \partial_+ C \partial_+
  C ,
\end{equation}
which is also non-unitary as the second term has the wrong sign. We will argue here
that these are two formulations of the same CFT, so that the $A, C$ system can
be viewed as a `bosonisation' of the $Q', P$ system.

\subsection {Bosonisation}

A standard
bosonisation  of the  $\beta \gamma$ system  takes it to the $\phi, \eta, \xi$ system, in the notation of e.g.\
 \cite{Polchinski:1998rq,Polchinski:1998rr}. A further bosonisation then takes it to a system of 2 scalars
$\phi, \chi$
\begin{equation}
  \beta \sim e^{- \phi + \chi}, \quad \gamma \sim e^{\phi - \chi} ,
\end{equation}
with
\begin{equation}
  T =  [\partial_+ \phi \partial_+ \phi - (1 - 2 \lambda')
  \partial^2_+ \phi] - [\partial_+ \chi \partial_+ \chi +
  \partial^2_+ \chi] .
\end{equation}
Note that $\phi$ has positive energy and $\chi$ has negative energy. The
central charge for $\phi$ is
\begin{equation}
  c_{\phi} = 3 (2 \lambda' - 1)^2 + 1 ,
\end{equation}
which for $\lambda' = 1$ gives
\begin{equation}
  c_{\phi} = 4 .
\end{equation}
while the central charge for $\chi$ is
\begin{equation}
  c_{\chi} = - 2 .
\end{equation}
Thus in total we get a CFT with central charge $c = 4 - 2 = 2$, as we should.

All OPEs, correlation functions and partition function agree so that $
(\beta, \gamma)$ and $(\phi, \chi)$ are regarded as giving 2 realisations of
the same $c = 2$ non-unitary chiral CFT. Part of the dictionary is:
\begin{equation}
  \delta (\gamma) \sim e^{- \phi} ,
\end{equation}
so that this can be regarded as an operator in both theories.

\subsection{A Different `Bosonisation'}

We now argue that the $(A, C)$ theory can be regarded as another realisation of
the same c=2 CFT. In other words, there is another bosonisation-type story for
the $Q', P$ system. First, note that the stress tensor is
\begin{equation}
  T_{+ +} = - Q'_+ \partial_+ P .
\end{equation}
Using
\begin{equation}
  F_+ =  \frac 1 2 (Q'_+ -  \partial_+ P) ,
\end{equation}
\begin{equation}
  G_+ = \frac 1 2 ( Q'_+ + \partial_+ P) ,
\end{equation}
$T_{+ +}$ can be written as
\begin{equation}
  T_{+ +} =  F_+ F_+ - G_+ G_+ .
\end{equation}

The introduction of the scalar $S$ can be viewed as a bosonisation, albeit of a rather trivial kind. On-shell $\partial _-Q'_+=0$ so that $dQ'=0$. Then there is a scalar  $S$ such that $Q'=d S$.
 Moreover, as $Q'$ is self-dual, $\partial_- S=0$ and $S$ is a chiral scalar, with
\begin{equation}
  Q_+' = \partial_+ S .
\end{equation}
The energy-momentum tensor is then
\begin{equation}
  T_{+ +} = - \partial_+S \partial_+ P .
\end{equation}
We can then use the chiral (holomorphic) fields $S, P$ with $ T_{+ +}$ to define a chiral
CFT. We define chiral fields
\begin{equation}
\label{acis}
  A =\frac 1 2( S-   P), \quad C = \frac 1 2(S+   P) ,
\end{equation}
so that
\begin{equation}
  T_{+ +} =\partial_+ A \partial_+ A - \partial_+ C \partial_+
  C .
\end{equation}
This gives the desired result: the relation between Sen's system with $g
= \bar{g} = \eta$ and the theory of chiral scalars $A, C$ is a bosonisation.
 
\section{Zero Modes}

One of the issues with bosonisation relations is that of missing zero modes. The bosonisation
\begin{equation}
  Q_+' = \partial_+ S ,
\end{equation}
relates the modes $S_n$ in the Laurent expansion of $S$ to the modes $Q'_n $
of $Q_+'$, apart from $S_0$. Indeed, $S$ only enters the Sen theory through
$Q_+' = \partial_+ S$ and so the constant part $S_0$ doesn't enter the theory.
Sen's theory gives $F_+ = \partial_+ A$ and not $A$ itself, so the theory doesn't give
the constant part of $A$.

There is a similar story for the bosonised $\beta \gamma$ system as the zero
mode of $\xi$ doesn't enter the theory -- see the discussion in \cite{Polchinski:1998rq,Polchinski:1998rr}.
There is a ``small formalism'' which doesn't include $\xi_0$ and a ``big
formalism'' which does. In fact, the usual formalism for amplitudes uses
$\xi_0,$ and this is related to pictures. (For example, the picture changing
operator is $[Q_{{BRS}}, \xi_0]$.) 
The formulation of superstring amplitudes requires extending the $\beta  \gamma$ system to include singular
functions of operators such as $\delta (\gamma)$, $\delta (\beta)$, and these
appear explicitly in amplitudes \cite{Polchinski:1998rq,Polchinski:1998rr}. The field $\xi$ itself arises as
the step function
\begin{equation}
  \xi = \theta (\beta) .
\end{equation}
In the same way, for the $Q'P$ CFT, there is a ``small formalism'' which doesn't
include $S_0$ and a ``big formalism'' which does, and there is a related question of whether to extend the $Q'P$ CFT to include singular functions of $P$ and $Q'$.

\noindent {\underline{Small formalism:}} In this formalism, there is a zero-mode $P_0 $
of $P$ but not a zero-mode $S_0$ of $S$. The zero modes (constant parts) of
$A, C$ are then
\begin{equation}
  A_0 = - \frac{1}{2} P_0 ,
\end{equation}
\begin{equation}
  C_0 = \frac{1}{2} P_0 .
\end{equation}
Then we can use $P_0$ to provide $A_0$ and constrain the zero-mode of $C$ to
be $C_0 = - A_0$. That way we get all the modes of $A$ from the formalism, but
constrain the zero-mode in the shadow sector.\footnote{Another approach discussed in \cite{Lambert:2023qgs} sets  $C_0=0$. This corresponds to fixing $S_0$ to be   $S_0=-P_0$ here.}

\noindent {\underline{Big formalism}}: In this formalism, there is both a zero-mode $P_0
$ of $P$ and a zero-mode $S_0$ of $S$. Then $A$ and $C$ have independent
zero modes. This requires enlarging the $Q'P$ CFT to include singular operators
such as $\delta (Q)$, $\theta (Q )$.

\section{Periodicity and Winding Modes}

A scalar field can be either single-valued, taking values in $\mathbb{R}$, or
periodic taking values in $S^1$, which allows for winding modes. 
The original theory is formulated in terms of the 1-form $Q'$ and the scalar $P$.
Suppose $P$ is periodic with
\begin{equation}
  P \sim P + 2 \pi R ,
\end{equation}
for some $R$.
Off-shell $dQ'$ need not vanish so there is no scalar $S$ and so no scalars $A,C$.
However, on-shell $Q'$ is closed and there is a chiral scalar $S$ such that
$Q'_+=\partial _+S$, and this can be used to define the chiral scalars 
\begin{equation}
\label{acisss}
  A =\frac 1 2( S-   P), \quad C = \frac 1 2(S+   P) .
\end{equation}
If $S$ is a non-compact boson, $S\in {\mathbb {R}}$, then $A$ and $C$ parameterise a cylinder
with $A-C$ periodic.
If $S$ is a compact boson with periodicity $S\sim S + 2\pi R'$ 
for some $R'$, then $A,C$ must  both have periodicity 
$\pi R$ and  periodicity $\pi R'$, 
which is only consistent with a toroidal geometry
if $R/R '$ is rational, i.e.
\begin{equation}
R= NR_0, \qquad R'=N' R_0
\end{equation}
for some radius $R_0$ and integers $N,N'$.\footnote{The case in which $R/R '$ is irrational can be interpreted as giving a non-commutative torus.}
It then follows that $A, C$ are periodic with period
$\pi R_0$,
\begin{equation}
A\sim A + \pi R_0, \qquad C\sim S+\pi R_0
\end{equation}
and $A,C$ take values in a square torus.

\section{OPEs, Vertex Operators and Correlation Functions}

The theory we are considering is a $\beta   \gamma$ system where $\beta=Q'$ has
weight 1 and $\gamma=P$ has weight zero. The Euclidean action is
\begin{equation}
  S = \int d^2 z \,Q' \bar{\partial}P .
\end{equation}
As
usual, for the cylinder with coordinates $\tau, \sigma$, the coordinates on
the Euclidean plane are $z = e^{\tau + i \sigma}$. On-shell, $Q' (z),
P (z)$ are holomorphic fields with the OPE
\begin{equation}
 Q'(z) P (w) \sim - \frac{1}{2\pi} \frac{1}{z - w} ,
\end{equation}
so that
\begin{equation}
  Q' (z) e^{i k P (w)} \sim -  \frac{1}{2\pi}
  \frac{1}{z - w} i k e^{i k P (w)} .
\end{equation}
The energy momentum tensor is
\begin{equation}
  T = - Q' \partial P .
\end{equation}
The bosonisation introduces a holomorphic field $S (z) $ with
\begin{equation}
  Q' = \partial S ,
\end{equation}
so that
\begin{equation}
  S (z) P (w) \sim -\frac{1}{2\pi} \log (z - w) ,
\end{equation}
and
\begin{equation}
  T = - \partial S \partial P .
\end{equation}
The field $S$ can be defined by an integral of $Q'(z)$
\begin{equation}
  S (z) = \int_{z_0}^z Q' ,
\end{equation}
on some contour from an arbitrary point $z_0$ to $z$, which is a non-local
operator in the $Q'   P$ system. Then $e^{i k S}$ is a
line-operator in the $Q'  P$  system
\begin{equation}
  e^{i k S (z)} = e^{i k \int^z Q'} .
\end{equation}
We will see below that with certain quantisation conditions  this operator is independent of the choice of contour.

Changing variables to $A,C$ defined by (\ref{acis}), the OPEs are
\begin{equation}
  C (z) C (w) \sim - \frac{1}{4\pi} \log (z - w), \quad A (z) A (w) \sim \frac{1}{4\pi} \log (z
  - w) ,
\end{equation}
and
\begin{equation}
  T =    \partial A \partial A -  \partial C  
  \partial C .
\end{equation}
Then the theory is one of a conventional scalar   $A$ and a
negative-energy non-unitary scalar $C$, with total central charge $c = 2$.

Vertex operators in the $AC$ system are made from regularised
(normal-ordered) combinations of $\partial A$, $\partial C$, $e^{i k
A}$, $e^{i p C}$ in the usual way. These can then be translated to
vertex operators and line operators in the $Q'P$ system. For
example,
\begin{equation}
  e^{i k A (z)} = \exp \left( \frac{i k}{ {2}} (S - P) (z)
  \right)  = \exp \left( \frac{i k}{ {2}} \left( \int^z Q' - P (z)
  \right) \right) = \exp \left( \frac{i k}{ {2}} \left( \int^z [Q' -
  \partial P] \right) \right) .
\end{equation}
and a correlation function of vertex operators
\begin{equation}
  \langle  e^{i k_1 A (z_1)} \ldots e^{i k_n A (z_n)} \rangle_{AC} ,
\end{equation}
in the $AC$ system can be calculated in the $Q' P$ system
as a correlation function of line operators
\begin{equation}
  \langle  e^{i k_1 L (z_1)} \ldots e^{i k_n L (z_n)} \rangle   ,\label{lincor}
\end{equation}
where
\begin{equation}
  L (z) = \frac{1}{{2}} \left( - P (z) + \int^z Q' \right) .
\end{equation}

We now turn to the dependence of the line operators and the correlation
functions on the choice of contours. 
Suppose the scalar $S$ is periodic, $S\sim S+2\pi R'$ for some $R'$, or, equivalently,  the
periods of $Q'$ on 1-cycles ${\mathcal{C}}$ are  quantised,
\begin{equation}
  \frac{1}{2 \pi R'} \oint_{\mathcal{C}} Q' \in \mathbb{Z} .
\end{equation}
We have seen that if $P$ is compactified on circle of radius $R$, then $R'=r_1R/r_2$ for some integers $r_1,r_2$.
Then for 2 different paths $\mathcal{P}, \mathcal{P}'$ from $z_0$ to $z$, the
combination $\mathcal{C}= \mathcal{P}'$-$\mathcal{P}$ defines a closed contour
and
\begin{equation}
  \int_{\mathcal{P}'} Q' - \int_{\mathcal{P}} Q' = \oint_{\mathcal{C}}
 Q' \in 2 \pi R'\mathbb{Z} ,
\end{equation}
so that the line operator
\begin{equation}
  e^{i k \int_{\mathcal{P}}  Q'} ,
\end{equation}
is independent of the choice of contour provided
\begin{equation}
  k = \frac{n}{R'} ,
\end{equation}
for some integer $n$. Then the correlation function (\ref{lincor}) is
well-defined provided $k_i = n_i /  R'$ for  integer $n_i$.
If the periods of $Q'$ were not quantised, or if the $k_i$ are not quantised as $k_i = n_i / R'$, then the line operators and their correlation functions could depend on the choice of contours.

\section{ The Gauged CFT and its  Symmetries}

\subsection{Gauged Actions}

A chiral CFT can be coupled to gravity by adding a Noether coupling to a
component $h_{- -}$ of the graviton
\begin{equation}
   \frac12\int  d^2 x\, h_{- -} T_{+ +} .
\end{equation}
For example, for a free scalar $A$, the gauged action is
\begin{equation}
  S = -\int  d^2 x\,(\partial_+ A \partial_- A - \tfrac12 h_{- -} \partial_+ A \partial_+ A)  .\end{equation}
This linear action is invariant under the gauge transformations
\begin{equation}
  \delta A = k_- \partial_+ A ,
\end{equation}
\begin{equation}
  \delta h_{- -} = 2\partial_- k_- +   k_-  \partial_+ h_{- -} -   h_{- -} 
  \partial_+ k_- ,
\end{equation}
for  a general $k_- (x)$ \cite{Hull:1988dp,Hull:1989wu}. Invariance under transformations with parameter $k_+
(x)$ would require coupling to a graviton $h_{+ +}$ and invariance under both
symmetries leads to an action with non-linear dependence on both $h_{+ +}$ and
$h_{- -}$ -- this is the usual gravitational coupling.

For the non-unitary theories considered here, the conventional  coupling to gravity
 $h_{- -} T_{+ +}+\dots$
 would give a
theory with negative energies. 
However, in the spirit of \cite{Hull:1989wu}, it is possible to
introduce a coupling to two gravitons, $h_{- -}$ and $\bar{h}_{- -}$ 
with 
$h_{- -} $ coupling to $\Theta_{++}=F_+ F_+$ and $\bar{h}_{- -}$  coupling to $\bar\Theta_{++}=-G_+ G_+$. 
This gives the linear coupling
\begin{equation}
\label{slin}
  S_{lin}=\frac12\int d^2 x\,(h_{- -} F_+ F_+ - \bar{h}_{- -} G_+ G_+)  .
\end{equation}
With $F, G$ given in terms of $Q', P$ by (\ref{FGQP}), this can be added to
the $Q', P$ action (\ref{QPact}) to give
\begin{equation}
\label{gag}
  S_{gauged}=\int
  d^2 x \, Q'_+ \partial_- P+  
  \frac12\int d^2 x\,(h_{- -} F_+ F_+ - \bar{h}_{- -} G_+ G_+)  .
\end{equation}
This can
be viewed as the action (\ref{act}) in the gauge in which
\begin{equation}
  \sqrt {-g} g^{\mu \nu} = \left( \begin{array}{cc}
 -   h_{- -} & 1\\
    1 & 0
  \end{array} \right), 
  \quad
   \sqrt {-\bar g}  \bar g^{\mu \nu} = \left( \begin{array}{cc}
  -   \bar h_{- -} & 1\\
    1 & 0
  \end{array} \right)
   .
\end{equation}
Setting $\bar{h}_{- -} = 0$ and adding coupling to $h_{+ +}$ then gives the
Sen action with non-linear dependence on the graviton. On the other hand,
keeping $\bar{h}_{- -}$ and adding coupling to $h_{+ +}$, $\bar{h}_{+ +}$
leads to the non-linear action (\ref{act}) of \cite{Hull:2023dgp}. It is remarkable that it is
possible to find a non-linear coupling to both gravitons. A geometric
formulation of this construction was given in \cite{Hull:2023dgp}.

\subsection{Gauge Symmetries} \label{gagsym}

In \cite{Hull:2023dgp}, it was shown that (\ref{act}) has two gauge symmetries, one with vector parameter $\zeta^\mu$ for which $g$ is the gauge field and one with 
vector parameter $\chi^\mu$ for which $\bar g$ is the gauge field.
This then implies that the gauged action (\ref{gag}) should have gauge symmetries with  vector parameter $\zeta_-$ for which $h$ is the gauge field, transforming as 
\begin{align}
	\delta h_{--}=2\partial _-\zeta_-+\dots ,
\end{align}
 and 
gauge symmetries with  vector parameter $\chi_-$ for which $\bar h$ is the gauge field, transforming as 
\begin{align}
\delta \bar h_{--}=2\partial _-\chi_-+\dots .
\end{align}
This is indeed the case. Using 
results from \cite{Hull:2023dgp,Hull:1988dp,Hull:1988si}, the action (\ref{gag})
with
\begin{equation}
  F_+ =    \frac 1 2 (Q'_+ -  \partial_+ P), \quad G_+ = \frac 1 2 ( Q'_+ + 
  \partial_+ P) ,  
\end{equation}
 is invariant under the $\zeta $ transformations
  \begin{eqnarray} 
    \delta h_{- -} & = &  2 \partial_- \zeta_- + \zeta_- \partial_+ h_{- -} -
    h_{- -} \partial_+ \zeta_-,\\
  \delta \bar h_{- -} &=&0 ,\nonumber
\\
\delta P&= &-\zeta_- F_+ ,\nonumber
\\
\delta Q'_+&= &\partial_+ (\zeta_- F_+) ,
\end{eqnarray}
so that
\begin{equation}
\delta G_+=0, \qquad \delta F_+=\partial_+(\zeta_- F_+)  .
\end{equation}
Under the $\chi$ transformations we have
\begin{eqnarray}
\delta  h_{- -} &=&0 ,\nonumber
\\
 \delta \bar{h}_{- -} & = & 2 \partial_- \chi_- + \chi_- \partial_+
     \bar{h}_{- -} - \bar{h}_{- -} \partial_+ \chi_- , \nonumber
     \\
\delta P&= &\chi_-G_+  ,\nonumber
\\
\delta Q'_+&=& \partial_+(\chi_-G_+)
\end{eqnarray}
so that
\begin{equation}
\delta G_+=\partial_+(\chi_-G_+), \qquad \delta F_+ = 0 .
\end{equation}
The diagonal subgroup with $ \zeta_-=\chi_-=\xi_-$ is the diffeomorphism symmetry,
with
\begin{equation}
\label{diffy}
\delta F_+= {\cal L}_\xi F_+, 
\quad
\delta G_+={\cal L}_\xi G_+,
\quad
\delta g_{\mu \nu}={\cal L}_\xi g_{\mu \nu},
\quad
\delta \bar g_{\mu \nu}={\cal L}_\xi \bar g_{\mu \nu},
\end{equation}
with all fields transforming with the Lie derivative ${\cal L}_\xi$.
See Appendix B for a derivation of these $\zeta,\chi,\xi$ transformations.

\subsection{Anomalies}

It is well-known that chiral CFTs suffer from gravitational anomalies, so that the two-dimensional diffeomorphism symmetry is anomalous.  These manifest themselves through a non-vanishing two-point function of the energy-momentum tensor. This in turn can be thought of as computing the one-point function of in the presence of a linearised perturbation $h_{\mu\nu}$: 
 \begin{align}
 \langle T_{\mu\nu} (x)\rangle_h &=  \int T_{\mu\nu}(x) e^{-iS+\tfrac{i}{2}\int h^{\lambda\rho} T_{\lambda\rho}  }	\nonumber \\
 & = \tfrac{i}{2}\int h^{\lambda\rho}(y)\langle T_{\mu\nu} (x) T_{\lambda\rho}(y)\rangle   dy +\dots .\end{align} 

In the case studied here, there are two independent 
symmetries of the action (\ref{gag}) with parameters $\zeta^\mu, \chi^\mu$ with corresponding 
 conserved currents $\Theta_{\mu\nu}, \bar\Theta_{\mu\nu}$ that enter the perturbation (\ref{slin}). Each of these is anomalous, and they combine to give the gravitational anomaly, which is the anomaly in the diagonal subgroup, the diffeomorphism symmetry acting as (\ref{diffy}).
It is a straight-forward exercise to compute, in either the $Q'P$ or $AC$ system,  that
\begin{align}
	\langle \Theta_{++}(z)\Theta_{++}(w)\rangle  	& =  \frac12 \frac1{4\pi^2 }\frac{1}{ (z-w)^4} , \nonumber
 \\
	 	\langle \Theta_{++}(z)\bar \Theta_{++}(w)\rangle  &=0 \nonumber , \\
	\langle \bar\Theta_{++}(z)\bar \Theta_{++}(w)\rangle 	 & =   \frac12\frac1{4\pi^2 }\frac{1}{ (z-w)^4},\nonumber\\
	\langle T_{++}(z)T_{++}(w)\rangle  	& =   \frac1{4\pi^2 }\frac{1}{ (z-w)^4} .
\end{align}
Thus we   find $c_{\Theta} = 1=c_{\bar\Theta} = 1$ and $c_{T}=2$, so that there is an anomaly in both the  $\zeta^\mu$ and $ \chi^\mu$ symmetries, which combine to give the anomaly in the diffeomorphism symmetry.
Furthermore there is no mixed anomaly between the two types of diffeomorphisms. This then implies that they can be separately cancelled  by coupling to suitable matter in both the physical and shadow sectors. That is, the anomaly in the $\zeta^\mu$ symmetry can be cancelled by adding further matter coupling to $h_{--}$ while the anomaly in the $\chi^\mu$ symmetry can be cancelled by adding further matter coupling to $\bar h_{--}$. This will be discussed further elsewhere.

\section{CFT In $d$-Dimensions}

Consider now the action (\ref{act}) in $d$ dimensions. In the special case in
which $g = \bar{g}$, we have $M=0$ and the action is
\begin{equation}
  \label{act2} S = \int \left(- \frac{1}{2} dP \wedge \ast dP +  Q \wedge dP
  \right) =    \int Q' \wedge dP ,
\end{equation}
where
\begin{equation}
  Q' = Q + \frac{1}{2} (dP + \ast dP) ,
\end{equation}
is self-dual, $Q' = \ast Q'$. 
This theory can be regarded as a higher dimensional analogue of the $\beta\gamma $ system.

The field equations are
\begin{equation}
  d Q' = 0 ,
\end{equation}
and
\begin{equation}
  d P = \ast d P .
\end{equation}
Then
\begin{equation}
  Q' = d S ,
\end{equation}
for some $q - 1$ form potential $S$ and
\begin{equation}
  d S= \ast d S .
\end{equation}
Then $P$ and $S$ are chiral $(q - 1)$-form gauge fields. The map from the $Q',
P$ system to the $S,P$ system can be thought of as a higher dimensional
version of the 2-dimensional \lq bosonisation' considered above. 

The self-dual field strengths (\ref{Giss}),(\ref{Fiss}) become
\begin{equation}
  \label{Gissa} G =    \frac 1 2 (dP +  {\ast} dP) +  \frac 1 2 Q = \frac 1 2 Q' +	 \frac{1}{4} (dP + \ast dP) ,
\end{equation}
and
\begin{equation}
  F =  \frac 1 2 Q  = \frac 1 2 Q' - \frac{1}{4} (dP + \ast dP) .
\end{equation}
Then $F=dA$, $G=dC$ with
\begin{equation}
  A =\frac 1 2( S-   P), \quad C = \frac 1 2(S+   P) .
\end{equation}

Again, it is
remarkable that this system can be coupled to two different gravitons. 
We introduce the   energy-momentum tensors $\Theta_{\mu\nu}(F)$ (\ref{Theta}) and  $\bar\Theta_{\mu\nu}(G)$  (\ref{barTheta})    and add the Noether coupling to two gravitons 
$\tilde h_{\mu\nu},  h_{\mu\nu}$
\begin{equation}
\label{Noeth}
S_{lin}=\frac 1 2 \int d^dx\,  [\tilde h^{\mu\nu}\bar \Theta_{\mu\nu}(G)+h^{\mu\nu}\Theta_{\mu\nu}(F)] ,
\end{equation}
generalising (\ref{slin}). This has a non-linear completion to a theory with two diffeomorphism-like symmetries, corresponding to the two gauge fields
$\tilde h_{\mu\nu},  h_{\mu\nu}$.
The action of \cite{Hull:2023dgp} for coupling to two metrics $\tilde g_{\mu\nu}$, $ g_{\mu\nu}$ can be expanded around flat space with $\tilde g_{\mu\nu}=\eta _{\mu\nu}+\tilde h_{\mu\nu}$, $  g_{\mu\nu}=\eta _{\mu\nu}+\tilde h_{\mu\nu}$, and to linear order in $\tilde h_{\mu\nu},  h_{\mu\nu}$ this corresponds to adding the coupling (\ref{Noeth}).

If the self-duality constraint on $Q'$ were dropped, the action $\int Q' \wedge dP$ would be a BF theory, which is a topological field theory  in the sense that it is independent of the metric and gives two flat gauge fields. Note also that  $\int dP\wedge *dP$ is a conformally invariant as $P $ is a $(d/2-1)$-form in $d$ dimensions, 
and  the self-duality condition  $Q'=\ast Q'$ is conformally invariant for such forms, so that this theory defines a CFT.  In particular the total energy momentum tensor  $T_{\mu\nu} = \Theta_{\mu\nu}+\bar\Theta_{\mu\nu}$ 
is traceless. Thus the theory of $A,C$ is also a CFT, However, with the self-duality constraint on $Q'$, the theory is of two self-dual gauge fields and is an interesting non-trivial conformal field theory in $d=4k+2$ dimensions that can be viewed as a higher-dimensional analogue of the holomorphic $\beta\gamma$ system in two dimensions.  The  bosonisation relates these two CFT's.

The self-dual $(q-1)$-form gauge fields $S,C$ naturally couple to $(q-1)$-form currents $j_{S}, j_C$ and to  $q-2$-branes which are referred to as self-dual as they have equal electric and magnetic charges, The currents $j_{S}, j_C$ can be thought of as the current densities of the corresponding branes.
It is not immediately obvious how these sources should couple to the $Q' P$ system. This was investigated in \cite{Hull:2025yww}, where it was argued that it was necessary to introduce a secondary  $(q-1)$-form current $J_{S}$ with
\begin{equation}
d^\dag J_{S}=j_{S} ,
\end{equation}
so that the coupling of $S$ to $j_{S}$ could be rewritten as the coupling of $Q'$ to $J_{S}$:
\begin{equation}
\int S \wedge \ast j_{S}=-\int Q' \wedge \ast J_{S} .
\end{equation}
For example, if $j_{S}$ is localised on some $(q - 1)$-dimensional surface ${\mathcal{S}}$, then $J_{S}$ is localised on a $q $-dimensional surface ${\mathcal{U}}$ whose boundary is ${\mathcal{S}}$, or includes ${\mathcal{S}}$.
The  $q$-dimensional surface ${\mathcal{U}}$ ending on the $(q-1)$-dimensional surface  ${\mathcal{S}}$ can be thought of as the world-volume ${\mathcal{U}}$ for a Dirac  $(q-1)$-brane ending on the 
world-volume $\mathcal{S}$ of a $(q-2)$-brane; see \cite{Hull:2025yww} for a detailed discussion.

In the $A,C$ theory, the natural local operators to consider are the generalisations of Wilson lines of the form
\begin{equation}
e^{i \int _{\mathcal{S}} A}, \qquad e^{i \int _{\mathcal{S}} C} ,
\end{equation}
for some $(q - 1)$-dimensional surface ${\mathcal{S}}$. The translation of these to the $Q',P$ theory follows from the above discussion.
The bosonisation relates these to  the following operators in the $Q',P$ theory:
\begin{equation}
e^{i \int _{\mathcal{S}} P}, \qquad e^{i \int _{\mathcal{U}} Q'} ,
\end{equation}
where the boundary of ${\mathcal{U}}$ includes ${\mathcal{S}}$.
The operator $e^{i \int _{\mathcal{U}} Q'}$ is the generalisation of the line operator considered earlier in two  dimensions.
As in that case, the operator is independent on the choice of surface ${\mathcal{U}}$ ending on ${\mathcal{S}}$ if the fluxes of $Q$ on $q$-cycles satisfy a quantisation condition.

\section{A Democratic Action}

A free self-dual gauge field is formulated in terms of a $d/2$-form field strength $F$ satisfying $F=\ast F$ and $d F=0$.
In any dimension $d$, a theory of a $(q-1)$-form gauge field  with a $q$ form field strength $F_{q}=d A$ has a dual formulation with dual $(d-q)$-form field strength $\tilde F_{d-q}=d\tilde  A$. The democratic formulation uses both field strengths $F_{q},\tilde F_{d-q}$ satisfying the equations
\begin{equation}
F_{q}=\ast \tilde F_{d-q}, \qquad d F_{q} =0, \qquad d \tilde F_{d-q} =0 .
\label{eqsa}
\end{equation}
Such formulations extend to certain interacting theories and give a duality covariant formulation of supergravities; see e.g.\ \cite{Cremmer:1997ct}. They give duality covariant equations of motion, but the construction of  an action 
runs into similar problems to those that arise for actions of self-dual gauge fields. For example, one can take the sum of the actions 
$\frac 1 2 \int F\wedge \ast F$ and $\frac 1 2 \int \tilde F\wedge \ast \tilde F$
for $F_{q}=d A$ and $\tilde F_{d-q}=d\tilde  A$ but then the equations of motion need to be supplemented by the constraint $F_{q}=\ast \tilde F_{d-q}$.

The Sen action and the action considered in this paper extend to this case with $Q$ and $P$ being taken to be sums of forms of different degree, giving the required equations (\ref{eqsa}), but at the expense of adding a shadow sector.
Mamade and  Zwiebach \cite{Mamade:2025vcp} have shown that the formulation of RR gauge fields in IIB supergravity that arises from superstring field theory are of precisely this type with $Q=Q_1+Q_3+Q_5+Q_7+Q_9$ a sum of forms of odd degree and $P=P_0+P_2+P_4+P_6+P_8$
a sum of forms of even degree. In particular, this confirms that the Sen action for the self-dual 4-form gauge field arises in superstring field theory through the Sen action.
For the IIA theory, $Q$ is a sum of forms of even degree and $P$ is a sum of forms of odd degree.

To see how this works, consider the action in $d$ dimensions for a pair of forms  
\begin{equation}
  \label{act23} S = \int \left(- \frac{1}{2} dP \wedge \ast dP +  Q \wedge dP
  \right) =    \int Q' \wedge dP ,
\end{equation}
where
\begin{equation}
  Q' = Q + \frac{1}{2} (dP + \ast dP) ,
\end{equation}
with   
\begin{equation}
  Q=Q_q+Q_{d-q}, \qquad P=P_{q-1}+P_{d-q-1}  ,
  \end{equation}
so that $Q'=Q'_q+Q'_{d-q}$ is a sum of a $q$-form and a $(d-q)$-form.
Then the action is
\begin{equation}
  \label{act24} S =     \int Q' \wedge dP=  \int [ Q'_q \wedge dP_{d-q-1}+ Q'_{d-q} \wedge dP_{d-q-1}] .
\end{equation}
Suppose we demand that $Q$ satisfy the duality relation
\begin{equation}
Q_{d-q}=\ast  Q_q 
\end{equation}
for $q < d/2$.
If $d=4k+2$, we can include self-duality for $Q_{d/2}$, while in $d=4k$ we can include two $d/2$ forms, $Q_{d/2}, \tilde Q_{d/2}$ with $\tilde Q_{d/2}=\ast Q_{d/2}$.
In dimensions in which $\ast^2=1$ when acting on $q$-forms, this  is the condition that $Q'$ is self-dual, $Q'=\ast Q'$ as $\ast Q_{d-q}=  Q_q$.
The field equations for $P_{q-1},P_{d-q-1}$
are
\begin{equation}
dQ'_q=0, \qquad d Q'_{d-q}=0 ,
\end{equation}
while the field equation for $Q'_q$ is
\begin{equation}
  \label{ergte}  dP_{q-1} =\ast dP_{d-q-1} .
\end{equation}
Then this gives a democratic action for two fields, one a gauge field with field strength $Q_q $ 
satisfying the field equations
\begin{equation}
d Q'_q=0,\qquad d ^\dag Q'_q=0 ,
\end{equation}
and another with potential $P_{q-1}$ satisfying the field equation
\begin{equation}
 d ^\dag d P_{q-1}=0 .
\end{equation}
This democratic formulation will be discussed further elsewhere.

\section{Conclusion}

In this paper we have explored the CFT that arises from the Sen formulation \cite{Sen:2015nph,Sen:2019qit}, generalised to include two metrics \cite{Hull:2023dgp}, in the case that the two metrics agree. This has the simple action
\begin{align}
S =  \int Q'\wedge dP	
\end{align}
where $Q'=\star Q'$ is a self-dual form in $d=4k+2$ dimensions. In two dimensions this is the well-known holomorphic $\beta\gamma$ system (where $\beta$ and $\gamma$ have scaling dimensions 1 and 0 respectively and   $c=2$). Without the self-duality constraint, we would have a topological $B F$ theory, but  by introducing self-duality we introduce a dependence on the metric while maintaining conformal invariance. We also argued that this theory has a natural bosonisation in terms of two self-dual fields but with opposite contributions to the energy momentum tensor and so is non-unitary. We also discussed how vertex operators in the bosonised theory can be mapped to line operators in the original $Q'P$ system. Furthermore we showed that this system can be coupled to two non-trivial metrics at the linearised level. Remarkably this bi-metric theory can then be resummed into a full non-linear theory with two metrics; namely the action constructed in \cite{Hull:2023dgp}.  Lastly, we introduced a generalization to an action involving  multi-forms which leads to a duality-symmetric democratic action.

 \bigskip\bigskip
\noindent{\bf\Large Acknowledgements}:
\bigskip

\noindent 
 The research of CH was supported by   the STFC Consolidated Grant    ST/X000575/1. N.L. is supported by the STFC grant ST/X000753/1.

\appendix
\section{A Trinity of Energy-Momentum Tensors}

In this Appendix, 
  we  compute the conserved currents $\Theta_{\mu\nu}$, $\bar \Theta_{\mu\nu}$ and $T_{\mu\nu}$ defined in subsection \ref{symcur}. First consider  $\Theta_{\mu\nu}	$. 
Note that $g$ only appears in the action via ${  M}$. For the choice $\bar g = \eta$,  $\Theta_{\mu\nu}$ was computed in \cite{Andriolo:2020ykk}. Let us recall that calculation here and extend it to a general choice of $\bar g$. 
To this end, we note that $(1-\star)(Q+{  M}(Q))=0$ and hence, since $\delta Q=0$ under a variation of the metric $g$,
\begin{align}
-(\delta \star) (Q+{ M}(Q)) +  (1-\star)\delta {   M}(Q)=0	.
\end{align}
where under $g\to g+\delta g$,  the change in the Hodge duality operator is $\star\to \star+( \delta\star)$.
Since self-dual-forms only have non-zero wedge products with anti-self-dual forms we find
\begin{align}
 (Q+{ M}(Q))\wedge (\delta \star) (Q+{   M}(Q)) = 2(Q+{  M}(Q)) \wedge \delta {  M}(Q)  .
\end{align}
Thus we have
\begin{align}
Q \wedge \delta {  M}(Q)  & =   (Q+{ M}(Q))\wedge \delta {  M}(Q)- {  M}(Q) \wedge \delta {  M}(Q)\\
 & =  \frac12(Q+{ M}(Q))\wedge (\delta \star) (Q+{ M}(Q))- { M}(Q) \wedge \delta { M}(Q) .
\end{align}
Next we note that $(1-\bar\star ){  M}=0$ and hence $(1-\bar\star )\delta {  M}=0$ since $\bar \star$ is not being varied. This in turn implies that ${  M}\wedge \delta {  M} =0$ since both are anti-self-dual  with respect to $\bar \star$. Therefore the last term vanishes and we find
\begin{align}
Q\wedge \delta {  M}	(Q)  =   2 F\wedge (\delta \star ) F ,
\end{align} 
where $F = \tfrac12 Q + \tfrac12 { M}(Q)=\star  F$. Thus we find the same energy momentum tensor that we would have found from varying $g$ in the action
\begin{align}
S_F &=     \frac12 \int   F\wedge \star  F \nonumber\\
& =  -\frac12\frac{1}{(2k+1)!}\int\sqrt{-g} F\cdot F , 
\end{align}
and only afterwards  imposing $F=\star  F$. The   energy momentum tensor is
\begin{align} 
\Theta_{\mu\nu} = \frac{1}{(2k)!} F_\mu\cdot F_\nu .
\end{align}
Here we have introduced the notation $X_\mu\cdot Y_\nu  = X_{\mu \lambda_1...\lambda_{2k}}Y_{\nu\rho_1...\rho_{2k}}  g^{\lambda_1\rho_1}...  g^{\lambda_{2k}\rho_{2k}}$ and $X \cdot Y  = X_{  \lambda_1...\lambda_{2k+1}}Y_{ \rho_1...\rho_{2k+1}}  g^{\lambda_1\rho_1}...\bar g^{\lambda_{2k+1}\rho_{2k+1}}$.
Note that the usual $g_{\mu\nu}F^2$ term vanishes as $F$ is an odd form and self-dual with respect to $\star $. It is now straight-forward to verify that $g^{\mu\nu}\Theta_{\mu\nu}=0$ and, on-shell,  $\nabla^\mu\Theta_{\mu\nu}=0$ as $dF=d\star F=0$.

Next we turn to the computation of $\bar \Theta_{\mu\nu}$. The dependence on $\bar g$  of the first   term  in the action is  standard. For the remaining two terms  we also need to know $\bar \delta Q$ and $\bar \delta {  M}$. We note that $(1-\bar\star)Q=0$ and hence
\begin{align}
-(\bar\delta\bar \star) Q + (1-\bar\star)\bar\delta Q =0	.
\end{align}
This implies that $(1+\bar\star)(\bar\delta\bar \star)  Q=0$ and hence $(\bar\delta\bar \star) Q=\tfrac12(1-\bar\star)(\bar\delta\bar \star) Q$. Thus we have
\begin{align}
-\frac12(1-\bar\star)(\bar\delta\bar \star) Q + (1-\bar\star)\bar\delta Q =0,	
\end{align}
which can be solved by
\begin{align}
\bar \delta Q = \frac12(\bar\delta\bar \star) Q  +  \Upsilon,
\end{align}
for any   $ \Upsilon $ that is self-dual: $\Upsilon=\tfrac12(1+\bar\star)\Upsilon$.

Next we need to compute $\bar\delta {  M}$. We observe that $(1+\bar\star)  {  M}=0$ and hence
\begin{align}
(\bar\delta\bar \star){  M} + (1+\bar\star)\bar\delta{ M}   =0	 .
\end{align}
Thus we have $(1-\bar\star)(\bar\delta\bar \star){ M}=0$ and hence 
\begin{align}
	\bar\delta{ M}=-\frac12 (\bar\delta\bar \star){  M}+ \Xi.
\end{align}
 for some $ \Xi$ that is anti-self-dual: $\Xi=\tfrac12(1-\bar\star)\Xi$. 
To continue, we note that 
\begin{align}
0&=(1-\star)(Q+{  M}(Q)) ,
\end{align}
which must be maintained when we vary $\bar g$. This leads to 
\begin{align}
0&=(1-\star)(\bar\delta Q+\bar\delta {  M}(Q)+{  M}(\bar \delta Q) )\nonumber\\
&=(1-\star)(\bar\delta Q+\bar\delta {  M}(Q) +M(\Upsilon))\nonumber\\
& =(1-\star)\left(\frac12(\bar\delta\bar \star)Q+\Upsilon -\frac12 (\bar\delta\bar \star){  M}(Q)+ \Xi(Q) +M(\Upsilon)\right) \nonumber\\
& =(1-\star)\left(-\frac12 (\bar\delta\bar \star){  M}(Q)+\frac12(\bar\delta\bar \star) Q + \Xi(Q) \right),
\end{align}
where in the second line we used the fact that ${ M}$ vanishes on anti-self-dual forms and in the fourth line that $ \Upsilon+M(\Upsilon)=\star(\Upsilon+M(\Upsilon))$. Thus we learn that
$-\frac12 (\bar\delta\bar \star){  M}(Q)+\frac12(\bar\delta\bar \star) Q + \Xi(Q)$ is also self-dual with respect to $\star$ and can be written as
\begin{align}
	-\frac12 (\bar\delta\bar \star){  M}(Q)+\frac12(\bar\delta\bar \star) Q + \Xi(Q) = \Pi + M(\Pi),
\end{align}
for some $\Pi$ which is self-dual with respect to $\bar\star$. 
Projecting on the self-dual and anti-self-dual parts we find
\begin{align}
\Pi = -\frac12 	(\bar\delta\bar \star){  M}(Q),\qquad M(\Pi) = \frac12(\bar\delta\bar \star) Q + \Xi(Q),
\end{align}
which allows us to identify
\begin{align}
	-\frac12 M(	(\bar\delta\bar \star){  M}(Q)) =\frac12(\bar\delta\bar \star) Q + \Xi(Q). 
\end{align}
From this we can determine $\Xi$:
\begin{align}
	 \Xi(Q)=-\frac12(\bar\delta\bar \star) Q-\frac12 M(	(\bar\delta\bar \star){  M}(Q)),
\end{align}
and hence  $\bar\delta M$:
\begin{align}
	  \bar\delta{ M}=-\frac12(\bar\delta\bar \star) Q-\frac12 (\bar\delta\bar \star){  M}-\frac12 M(	(\bar\delta\bar \star){  M}(Q)).
\end{align}
A similar expression  also appeared in \cite{Lambert:2019diy}
 for the case $\bar g = \eta$. 

Let us return to the variation of the action and note that 
 \begin{align}
 \bar\delta \left(Q\wedge { M}(Q)\right)	 &= 2\bar\delta Q\wedge  {  M}(Q) + Q\wedge \bar\delta{  M}(Q)\nonumber \\
 & = 2\Upsilon\wedge M(Q) + Q\wedge\Xi 
 \end{align}
where we have used the fact that (anti-)self-dual forms  have vanishing wedge products with themselves. Thus  we find
\begin{align}\label{QQ}
 \bar\delta \left(Q\wedge {  M}(Q)\right)	 
 & =  2\Upsilon\wedge M(Q) -\frac12 Q\wedge (\bar\delta\bar \star) Q - \frac12Q\wedge M(	(\bar\delta\bar \star){  M}(Q)) \nonumber\\
 & =   2\Upsilon\wedge M(Q)-\frac12 Q\wedge (\bar\delta\bar \star) Q  - \frac12(\bar\delta\bar \star){  M}(Q)\wedge M(Q)  .
 \end{align}
 
 Finally  we need to determine  $\Upsilon$. To do this, we first look at  the case $\Upsilon=0$ and drop the last  term in  (\ref{QQ}).
Similarly to the case above,  this is leads to the same energy-momentum tensor that we would get from 
 \begin{align}
S' &=   -\int \frac12  dP\wedge\bar\star dP- \frac12( Q+\bar\star Q) \wedge dP+ \frac18 Q\wedge \bar \star  Q ,
\end{align}
by varying $\bar g_{\mu\nu}$ and only then imposing $Q= \bar \star Q$. This gives
\begin{align}
\bar \Theta_{\mu\nu}=-\frac{1}{(2k)!}&\left[ (dP)_\mu\bar \cdot  (dP)_\nu +\frac12 Q_\mu\bar\cdot (dP)_\nu+\frac12 Q_\nu \bar\cdot (dP)_\mu+ \frac14 Q_\mu\bar\cdot Q_\nu \right]\nonumber\\
& +\frac{1}{2(2k+1)!}\bar g_{\mu\nu}\left[ (dP) \bar\cdot  (dP)  +  Q \bar\cdot (dP)   \right] +\bar\Theta'_{\mu\nu},
\end{align} 
where $\bar\Theta'_{\mu\nu}$ is the contribution arising from $\Upsilon$ and the last term in (\ref{QQ}). Note that now $X_\mu\bar \cdot Y_\nu  = X_{\mu \lambda_1...\lambda_{2k}}Y_{\nu\rho_1...\rho_{2k}}\bar g^{\lambda_1\rho_1}...\bar g^{\lambda_{2k}\rho_{2k}}$ and $X \bar\cdot Y  = X_{  \lambda_1...\lambda_{2k+1}}Y_{ \rho_1...\rho_{2k+1}}\bar g^{\lambda_1\rho_1}...\bar g^{\lambda_{2k+1}\rho_{2k+1}}$.
 A little calculation shows that  this can be written as 
 \begin{align}
\bar \Theta_{\mu\nu}=-\frac{1}{(2k)!} G_\mu\bar\cdot G_\nu +\frac14 \frac{1}{(2k)!}(dP-\bar\star dP)_\mu\bar\cdot (dP-\bar\star dP)_\nu +\bar\Theta'_{\mu\nu}.
\end{align}
However the second term is not conserved on-shell. 

To fix this we need to consider  $\Upsilon\ne0 $ and include the last term in (\ref{QQ}). This  gives an additional variation
\begin{align}
\delta' S &= \int \frac12\Upsilon\wedge (dP-\bar \star dP)+  \frac12 \Upsilon \wedge M(Q) - \frac18(\bar\delta\bar \star){  M}(Q)\wedge M(Q) .
\end{align}
A natural choice is $ \Upsilon = \tfrac12(\bar\delta\bar \star) M(Q)$, which we have seen is self-dual, leading to 
\begin{align}
\bar \delta Q &= \frac12	(\bar\delta\bar \star) Q  +  \frac12\bar\delta \bar \star M(Q) \nonumber\\
& =(\bar\delta\bar \star)F,
\end{align}	
and 
\begin{align}
\delta'S &= - \int \frac14(dP-\bar \star dP)\wedge (\bar\delta\bar \star)M(Q)+  \frac18 M(Q) \wedge (\bar\delta\bar \star) M(Q) .
\end{align}
Following similar arguments to those used above this leads to the addition contribution\begin{align}
\bar\Theta'_{\mu\nu} = -\frac14\frac{1}{(2k)!} \left( (dP-\bar\star dP)_\mu\bar\cdot M(Q)_\nu + (dP-\bar\star dP)_\nu\bar\cdot M(Q)_\mu +    M_\mu(Q)\bar\cdot M_\nu(Q)\right)  ,
\end{align}
and hence we find the full shadow energy-momentum tensor is
\begin{align} 
\bar \Theta_{\mu\nu}=-\frac{1}{(2k)!} G_\mu\bar\cdot G_\nu -\frac14 \frac{1}{(2k)!}(dP-\bar\star dP+M(Q))_\mu\bar\cdot (dP-\bar\star dP+M(Q))_\nu .	
\end{align}
Here we see that the first term is conserved on-shell, i.e. it's divergence with respect to $\bar\nabla$ vanishes on-shell,  whereas the second term vanishes on-shell. As a result,  $\bar \nabla^\mu \bar \Theta_{\mu\nu}=0$ on-shell.
 
Lastly we consider an infinitesimal  diffeomorphism $x^\mu \to  x^\mu +\xi^\mu$ so that
\begin{align}
\bar g^{\mu\nu}&\to 	\bar g^{\mu\nu} - \bar \nabla_\lambda \xi^\mu \bar  \bar g^{\lambda\nu} -  \bar \nabla_\lambda\xi^\nu   \bar g^{\mu\lambda}\nonumber\\
 g^{\mu\nu}&\to 	  g^{\mu\nu} -\nabla_\lambda   \xi^\mu   g^{\lambda\nu} -  \nabla_\lambda  \xi^\nu g^{\mu\lambda} .
\end{align}
The action  is invariant under such diffeomorphisms, up to possible boundary terms which we will discard.  We observe that 
\begin{align}
\delta S & = \int 	\sqrt{-\bar g}\bar \Theta_{\mu\nu}\bar \nabla_\lambda \xi^\mu\bar g^{\lambda\nu} +  \int 	 \sqrt{-  g}\Theta_{\mu\nu} \nabla_\lambda \xi^\mu  g^{\lambda\nu} \nonumber\\
&\cong - \int \left(	\sqrt{-\bar g}\bar \nabla_\lambda  \bar\Theta_\mu{}^\lambda  +  \sqrt{- g}\nabla_\lambda   \Theta_\mu{}^\lambda\right) \xi^\mu ,
\end{align}
where $\bar \Theta_{\mu}{}^\lambda =\bar \Theta_{\mu\nu}\bar g^{\nu\lambda} $ and $ \Theta_{\mu}{}^\lambda =  \Theta_{\mu\nu}g^{\nu\lambda} $.
  From the condition $\delta S =0$ we deduce that
  \begin{align}
\sqrt{-\bar g}\bar \nabla_\lambda  \bar\Theta_\mu{}^\lambda  + \sqrt{-  g} \nabla_\lambda   \Theta_\mu{}^\lambda =0	 .
\end{align}
However we have seen that each of these two terms vanishes independently.  This is to be expected from the discussion in \cite{Hull:2023dgp} which shows that there are two independent vector-generated gauge transformations which act separately on $g$ and $\bar g$ whereas diffeomorphisms arising from coordinate transformations correspond to the diagonal subgroup. However when $g=\bar g$ it is natural to identify a third energy-momentum tensor
\begin{align}
T_{\mu\nu} & =  \Theta_{\mu\nu}+\bar \Theta_{\mu\nu}\nonumber\\
& \approx 	\frac{1}{(2k)!} F_\mu \cdot F_\nu -\frac{1}{(2k)!} G_\mu \cdot G_\nu . 
\end{align}
  whose conservation follows from diffeomorphism invariance.

\section{Derivation of the gauge transformations}

This appendix gives the derivation of the gauge symmetries presented in section \ref{gagsym}. 
The gauged action is
\begin{equation}
  S_{gauged} = \int d^2 x \hspace{0.17em} Q'_+ \partial_- P +\frac{1}{2} 
  \int d^2 x \hspace{0.17em} (h_{- -} F_+ F_+ - \bar{h}_{- -} G_+ G_+) ,
\end{equation}
where
\begin{equation}
  F_+ = \frac{1}{2}  (Q'_+ - \partial_+ P), \quad G_+ = \frac{1}{2}  (Q'_+ +
  \partial_+ P) .
\end{equation}

Consider the $ \zeta$-transformations
\begin{eqnarray}
  \delta \bar{h}_{- -} & = & 0 \\
  \delta P & = & - \zeta_- F_+ \\
  \delta Q'_+ & = & \partial_+  (\zeta_- F_+) 
\end{eqnarray}
so that
\begin{equation}
  \delta G_+ = 0, \qquad \delta F_+ = \zeta_- \partial_+ F_+ + F_+ \partial_+
  \zeta_- .
\end{equation}
Then the variation of the kinetic term is

\begin{eqnarray}
  \delta \int d^2 x \hspace{0.17em} Q'_+ \partial_- P & = & \int d^2 x \left[ 
  \hspace{0.17em} \partial_+ (\zeta_- F_+) \partial_- P - Q'_+ \partial_- 
  (\zeta_- F_+) \right] \nonumber\\
  & = & \int d^2 x \zeta_- F_+ [- \partial_+ \partial_- P + \partial_- Q'_+]
  \nonumber\\
  & = & \int d^2 x \zeta_- F_+ \partial_- [- \partial_+ P + Q'_+] \nonumber\\
  & = & 2 \int d^2 x \zeta_- F_+ \partial_- F_+ \nonumber\\
  & = & - \int d^2 x (\partial_- \zeta_-) F_+ F_+ .
\end{eqnarray}
This can be cancelled by choosing
\begin{equation}
  \begin{array}{lll}
    \delta h_{- -} & = &  2 \partial_- \zeta_- + X_{- -} ,
  \end{array}
\end{equation}
for some $X$. Then
\begin{eqnarray}
  \delta S_{gauged} & = & 
  \frac{1}{2}  \int d^2 x \hspace{0.17em} (X_{- -}
  F_+ F_+ + 2 h_{- -} F_+ \delta F_+) \nonumber\\
  & = & 
   \frac{1}{2}  \int d^2 x \hspace{0.17em} (X_{- -} F_+ F_+ + 2 h_{-
  -} F_+ \partial_+ (\zeta_- F_+)) \nonumber\\
  & = & 
  \frac{1}{2}  \int d^2 x \hspace{0.17em} (X_{- -} F_+ F_+ + 2 h_{-
  -}  (\partial_+ \zeta_-) F_+ F_+ + h_{- -} \zeta_- \partial_+ [F_+]^2)
  \nonumber\\
  & = & 
  \frac{1}{2}  \int d^2 x \,  F_+ F_+ \hspace{0.17em} (X_{- -}
  + h_{- -}  (\partial_+ \zeta_-) - \zeta_- \partial_+ h_{- -})  .
\end{eqnarray}
This will vanish if
\begin{equation}
  X_{- -} = - h_{- -} \partial_+ \zeta_- + \zeta_- \partial_+ h_{- -} ,
\end{equation}
so that
\begin{equation}
  \begin{array}{lll}
    \delta h_{- -} & = &  2 \partial_- \zeta_- + \zeta_- \partial_+ h_{- -} -
    h_{- -} \partial_+ \zeta_- .
  \end{array}
\end{equation}

Consider next the $\chi$ transformations
\begin{eqnarray}
  \delta h_{- -} & = & 0 \\
  \delta P & = & \chi_- G_+ \\
  \delta Q'_+ & = & \partial_+  (\chi_- G_+) .
\end{eqnarray}
so that
\begin{equation}
  \delta F_+ = 0, \qquad \delta G_+ = \chi_- \partial_+ G_+ + G_+ \partial_+
  \chi_- .
\end{equation}
Then
\begin{eqnarray}
  \delta \int d^2 x \hspace{0.17em} Q'_+ \partial_- P & = & \int d^2 x \left[ 
  \hspace{0.17em} \partial_+ (\chi_- G_+) \partial_- P + Q'_+ \partial_- 
  (\chi_- G_+) \right] \nonumber\\
  & = & \int d^2 x \chi_- G_+ [- \partial_+ \partial_- P - \partial_- Q'_+]
  \nonumber\\
  & = & - \int d^2 x \chi_- G_+ \partial_- [\partial_+ P + Q'_+] \nonumber\\
  & = & - 2 \int d^2 x \chi_- G_+ \partial_- G_+ \nonumber\\
  & = & \int d^2 x (\partial_- \chi_-) G_+ G_+  .
\end{eqnarray}
These can be canceled by choosing
\begin{equation}
  \begin{array}{lll}
    \delta \bar{h}_{- -} & = & 2 \partial_- \chi_- + Y_{- -} ,
  \end{array}
\end{equation}
for some $Y$. Then
\begin{eqnarray}
  \delta S_{gauged} & = & -\frac{1}{2}  \int d^2 x \hspace{0.17em} (Y_{- -} G_+ 
  G_+ + 2 \bar{h}_{- -} G_+ \delta G_+)\nonumber \\
  & = & -\frac{1}{2}  \int d^2 x \hspace{0.17em} (Y_{- -} G_+ G_+ + 2
  \bar{h}_{- -} G_+ \partial_+ (\chi_- G_+))
  \nonumber
  \\
  & = & -\frac{1}{2}  \int d^2 x \hspace{0.17em} (Y_{- -} F_+ F_+ + 2
  \bar{h}_{- -}  (\partial_+ \chi_-) G_+ G_+ + \bar{h}_{- -} \chi_- \partial_+
  [G_+]^2)
  \nonumber
  \\
  & = & -\frac{1}{2}  \int d^2 x \, G_+ G_+ \hspace{0.17em} (Y_{- -} +
  \bar{h}_{- -}  (\partial_+ \chi_-) - \chi_- \partial_+ \bar{h}_{- -}) .
\end{eqnarray}
This will vanish if
\[ Y_{- -} = \chi_- \partial_+ \bar{h}_{- -} - \bar{h}_{- -}  (\partial_+
   \chi_-) , \]
leading to
\[ \begin{array}{lll}
     \delta \bar{h}_{- -} & = &  2 \partial_- \chi_- + \chi_- \partial_+
     \bar{h}_{- -} - \bar{h}_{- -} \partial_+ \chi_- .
   \end{array} \]

 Consider now the diagonal subgroup of these two symmetries where $\chi_- =
\zeta_- = \xi_-$. This gives a symmetry
\[ \begin{array}{lll}
     \delta h_{- -} & = &  2 \partial_- \xi_- + \xi_- \partial_+ h_{- -} - h_{-
     -} \partial_+ \xi_-,\nonumber\\ 
     \delta \bar{h}_{- -} & = & 2 \partial_- \xi_- + \xi_- \partial_+
     \bar{h}_{- -} - \bar{h}_{- -} \partial_+ \xi_- ,
   \end{array} \]
\[      \delta P  = \xi_- (G_+ - F_+) = \xi_- \partial_+ P ,\]
\[ \delta Q'_+ = \partial_+  (\xi_-  (F_+ + G_+)) = \partial_+  (\xi_- Q'_+) ,
\]
so that
\[ \delta F_+ =  \partial_+  (\xi_- F_+) ,\quad \delta G_+ =  \partial_+  (\xi_- G_+) .
\]

Let us now compare this to what we expect from diffeomorphisms. 
The transformation of a 1-form $V$ under an infinitesimal diffeomorphism generated by a
vector field $\xi$ 
is given by the Lie derivative
\begin{align}  
\delta V_\mu=
{\cal L}_\xi V_{\mu} & = \xi^\lambda \partial_\lambda V_{\mu} + \partial_{\mu}\xi^{\nu}V_{\nu}
.
\end{align} 
If $V$ and $\xi$ are self-dual so that $V_-=0$, $\xi^-=0$, then the transformation becomes
\begin{equation}
\delta V_+=\xi^+ \partial_+ V_{+} + V_{+}\partial_{+}\xi^{+}= \partial_+ (V_{+} \xi_-) ,
\end{equation}
 where $\xi_-=\xi^+$.
 Thus the transformations of $F_+,G_+,Q'_+$ are all through the Lie derivative.
 
 Note that $h_{--}$ and $\bar h_{--}$ actually appear in the action as the $++$ components of   the inverse metrics densities
\begin{equation}
  \sqrt {-g} g^{\mu \nu} = \left( \begin{array}{cc}
  -  h_{- -} & 1\\
    1 & 0
  \end{array} \right), 
  \quad
   \sqrt {-\bar g}  \bar g^{\mu \nu} = \left( \begin{array}{cc}
   -  \bar h_{- -} & 1\\
    1 & 0
  \end{array} \right)
   . 
\end{equation}
The variation of $h_{--}$ and $\bar h_{--}$ can be read off  from the variation of $ \sqrt {-  g}   g^{++}$ and $ \sqrt {-\bar g}  \bar g^{++}$ respectively. The transformation
\begin{equation}
\delta g_{\mu\nu}={\cal L}_\xi g_{\mu\nu} ,
\end{equation}
gives, for a self-dual parameter self-dual   $\xi^-=0$,
\begin{equation}
\delta h_{- -}  =   2 \partial_- \xi_- + \xi_- \partial_+ h_{- -} - h_{-
     -} \partial_+ \xi_- ,
\end{equation}
where $\xi_-=\xi^+$.
  Similarly, 
\begin{equation}
\delta \bar g_{\mu\nu}={\cal L}_\xi \bar g_{\mu\nu} ,
\end{equation}
gives
\begin{equation}
 \delta \bar{h}_{- -} =  2 \partial_- \xi_- + \xi_- \partial_+
     \bar{h}_{- -} - \bar{h}_{- -} \partial_+ \xi_- .
\end{equation}
Moreover, with $\xi^-=0$,   the remaining components of $\delta \left( \sqrt{-g} g^{\mu\nu}\right)$ and $ \delta \left( \sqrt{-\bar g} \bar g^{\mu\nu}\right)$ are zero. 
In this way, we obtain
  the diffeomorphism  symmetry (\ref{diffy}).

\end{document}